
\overfullrule = 0pt
\magnification = 1200
\baselineskip 21pt
\def\tr{\mathop{\rm tr}\nolimits}

\centerline{\bf ADDENDUM}

\bigskip
\bigskip

\centerline{\bf INTEGRABILITY OF OPEN SPIN CHAINS WITH}
\centerline{\bf QUANTUM ALGEBRA SYMMETRY}

\bigskip

\centerline{[INT. J. MOD. PHYS. A, Vol. 6, 5231 (1991)]}

\bigskip

\medskip

\centerline{Luca Mezincescu and Rafael I. Nepomechie}
\centerline{Department of Physics}
\centerline{University of Miami, Coral Gables, FL 33124}

\bigskip
\bigskip

In the above reference, we prove that certain quantum spin chains
with quantum-algebra symmetry are integrable. Specifically, these
are the quantum-algebra-invariant open chains associated with the affine
Lie algebras
$A^{(1)}_1$, $A^{(2)}_{2n}$, $A^{(2)}_{2n - 1}$, $B^{(1)}_{n}$, $C^{(1)}_{n}$,
and $D^{(1)}_{n}$ in the fundamental representation. Conspicuously
absent from this list is $A^{(1)}_n$ for $n>1$. This is because in order
to demonstrate integrability, we assume that the corresponding $R$ matrix has
crossing symmetry, which is not true in the case $A^{(1)}_n$ for $n>1$.

Sklyanin has demonstrated${}^1$ that an integrable open spin chain
can be constructed with an $R$ matrix that satisfies nothing more than
the Yang-Baxter equation. In particular, no assumption of crossing
symmetry is necessary. This suggests that the quantum-algebra-invariant
open chain associated with $A^{(1)}_n$ for $n>1$ may be integrable.
In this Addendum, we show that this is in fact the case.

Let $R(u)$ be a solution of the Yang-Baxter equation corresponding to a
non-exceptional affine Lie algebra $g^{(k)}$ in the fundamental representation.
According to Bazhanov${}^2$ and Jimbo${}^3$, such an $R$ matrix has
$PT$ symmetry
$${\cal P}_{12}\ R_{12}(u)\ {\cal P}_{12} \equiv R_{21}(u)
= R_{12}(u)^{t_1 t_2} \,,   \eqno(1)$$
and is unitary
$$R_{12}(u)\  R_{21}(-u) = \zeta (u) \,,  \eqno(2) $$
where $\zeta (u)$ is some even scalar function of $u$.
Moreover, according to Reshetikhin and Semenov-Tian-Shansky${}^4$, there exists
a matrix $M$ such that
$$\{\{\{R_{12}(u)^{t_2}\}^{ -1}\}^{t_2}\}^{-1}
= {\zeta(u+\rho)\over \zeta(u + 2\rho)}
M_2\ R_{12}(u + 2\rho) M_2^{-1} \,. \eqno(3) $$
Also, $M^t = M$ and
$$ \left[ R(u) \,, M \otimes M \right] = 0 \,. \eqno(4) $$
We do {\it not} assume that $R$ has crossing symmetry.

Let us introduce matrices $K^-(u)$ and $K^+(u)$ as before${}^5$:
$$ R_{12}(u - v)\ K^-_1(u)\ R_{21} (u + v)\ K^-_2(v)
= K^-_2(v)\ R_{12}(u + v)\ K^-_1(u)\ R_{21}(u - v) \eqno(5) $$
and
$$\eqalignno{
&R_{12}(-u + v)\ K^+_1(u)^{t_1}\ M^{-1}_1\ R_{21} (-u -v -2\rho)\
M_1\ K^+_2(v)^{t_2} \cr
& \quad = K^+_2(v)^{t_2}\ M_1\ R_{12}(-u - v- 2\rho)\ M^{-1}_1\ K^+_1(u)^{t_1}\
R_{21}(-u +v)  \,. &(6) \cr}$$
Sklyanin${}^1$ has found a more general set of relations, which reduces
to this set when $R$ satisfies Eqs. (1) and (2).

With the help of Eqs. (3) and (4), one can verify that
$$K^+(u) = K^-(-u-\rho)^t M \eqno(7) $$
is an automorphism. For all cases except $D^{(2)}_n$ (so, in particular,
for the case $A^{(1)}_n$),
$$K^-(u) = 1 \eqno(8) $$
is a solution of Eq. (5). Hence,
$$K^+(u) = M \eqno(9) $$
is a solution of Eq. (6).
And hence, the corresponding commuting transfer matrix is${}^{6,5}$
$$t(u) = \tr K^+(u)\ T(u)\ K^-(u)\ T(-u)^{-1}
       = \tr M\ T(u)\ T(-u)^{-1} \,. \eqno(10)$$

This transfer matrix commutes also with the generators of the quantum algebra
$U_q[g_0]$, where $g_0$ is the maximal finite-dimensional subalgebra of
$g^{(k)}$. Indeed, let us recall that the monodromy matrix $T(u)$, upon
suitably taking the limit $u \rightarrow \pm \infty$, becomes an upper or
lower triangular matrix $T_\pm$ whose entries are elements of $U_q[g_0]$.
The proof of the formula
$$ \left[ T_\pm \,, t(u) \right] = 0  \eqno(11) $$
is essentially the one given (following Kulish and Sklyanin${}^7$)
in Ref. 8.

In conclusion, the case $A^{(1)}_n$ can be treated in almost the same way
as the other cases which we have previously studied. The only difference is
that the crossing relation for the $R$ matrix is replaced by the weaker
relation (3).

A number of interesting questions remain:
\item{1.} One may try to extend this analysis to the case $D^{(2)}_n$, for
which
Eq. (8) does not hold. Moreover, the exceptional cases have not yet been
investigated.
\item{2.} One may try to solve the $A^{(1)}_n$ chains by the analytical Bethe
Ansatz. This would require an extension of the known${}^{9,10}$ formalism,
which assumes crossing symmetry.
\item{3.} One may further explore${}^1$ open chains associated with
a matrix $R(u)$ which does not have $PT$ symmetry, or with a matrix
$R(u,v)$ which does not depend on the difference $u-v$. The chiral Potts model
is presumably an example of the second type.

\bigskip

We are indebted to E.K. Sklyanin for sharing with us his unpublished
work${}^1$. We also thank N. Yu. Reshetikhin for a discussion on the scalar
factor in Eq. (3). This work was supported in part by the National Science
Foundation under Grant No. PHY-90 07517.

\vfill\eject

\noindent
{\bf References}

\medskip

\item{1.}E.K. Sklyanin, unpublished.

\item{2.}V.V. Bazhanov, Phys. Lett. {\it 159B} (1985) 321;
Commun. Math. Phys. {\it 113} (1987) 471.

\item{3.}M. Jimbo, Commun. Math. Phys. {\it 102} (1986) 537;
{\it Lecture Notes in Physics}, Vol. 246 (Springer, 1986) 335.

\item{4.}N. Yu. Reshetikhin and M.A. Semenov-Tian-Shansky, Lett. Math. Phys.
{\it 19} (1990) 133. See also I. Frenkel and N. Yu. Reshetikhin,
preprint.

\item{5.}L. Mezincescu and R.I. Nepomechie, J. Phys. {\it A24} (1991) L17.

\item{6.}E.K. Sklyanin, J. Phys. {\it A21} (1988) 2375.

\item{7.}P.P. Kulish and E.K. Sklyanin, J. Phys. {\it A24} (1991) L435.

\item{8.}L. Mezincescu and R.I. Nepomechie, Mod. Phys. Lett. {\it A6} (1991)
2497.

\item{9.}N. Yu. Reshetikhin, Lett. Math. Phys. {\it 14} (1987) 235.

\item{10.}L. Mezincescu and R.I. Nepomechie, Nucl. Phys. {\it B372} (1992) 597.

\end